\documentclass[aps,prd,reprint,superscriptaddress,nofootinbib,floatfix]{revtex4-1}

\usepackage{amsfonts,amssymb,amsmath}
\usepackage{epsfig}
\usepackage{bm}
\usepackage{slashed}
\usepackage{latexsym}
\usepackage{natbib}
\usepackage{url}
\usepackage{dcolumn}
\usepackage{color}
\usepackage{graphicx}
\usepackage{psfrag}
\usepackage{ulem}
\usepackage{hyperref}
\hypersetup{colorlinks   = true,
            urlcolor     = blue,
            citecolor    = blue,
            linkcolor    = blue,
            menucolor    = blue,
            anchorcolor  = blue,
            filecolor    = blue
}
\everymath{\displaystyle}
\renewcommand{\arraystretch}{1.5}

\begin{document}

\title{Impact of the cosmic neutrino background on black hole superradiance}

\newcommand{\UniSA}{\affiliation{Dipartimento di Fisica ``E.R.\ Caianiello'', Universit\`a degli Studi di Salerno,\\ Via Giovanni Paolo II, 132 - 84084 Fisciano (SA), Italy}}
\newcommand{\INFN}{\affiliation{Istituto Nazionale di Fisica Nucleare - Gruppo Collegato di Salerno - Sezione di Napoli,\\ Via Giovanni Paolo II, 132 - 84084 Fisciano (SA), Italy}}

\author{Gaetano Lambiase}
\email{lambiase@sa.infn.it}
\UniSA \INFN
\author{Tanmay Kumar Poddar}
\email{poddar@sa.infn.it}
\INFN
\author{Luca Visinelli}
\email{lvisinelli@unisa.it}
\UniSA \INFN
 
\begin{abstract}
We assess the effect of the Cosmic Neutrino Background (C$\nu$B) on superradiant instabilities caused by an ultralight scalar field around spinning black holes (BHs). When the scalar couples to neutrinos via a Yukawa interaction, thermal corrections from the C$\nu$B induce a quartic self-interaction and an effective mass term for the scalar. We show that, for Yukawa couplings as small as $y_{\phi \nu} \sim 10^{-16}$ (for astrophysical BHs) or $10^{-20}$ (for supermassive BHs), the quartic term can quench the instability and set observable bounds, even if the scalar does not constitute dark matter. We assess the robustness of these constraints against several sources of uncertainty, including gravitational focusing of relic neutrinos, galactic clustering, and non-linear backreaction. An enhanced local neutrino density weakens the bounds by up to an order of magnitude compared to a uniform background, yet the induced self-interaction remains strong enough to significantly affect the superradiant dynamics. Our results open a new observational window on neutrino-coupled scalars via BH superradiance.
\end{abstract}

\date{\today}
\maketitle

\section{Introduction}
\label{sec:introduction}

Rotating black holes (BHs) provide an excellent laboratory for exploring ultralight particles through the process of particle superradiance~\cite{Penrose:1969pc, Penrose:1971uk, Press:1972zz, Zouros:1979iw, Detweiler:1980uk, Gaina:1988nf}. In this phenomenon, a bosonic field of mass $m_\phi$ forms a bound state around a BH of mass $M_{\rm BH}$, with its occupation number amplified when the BH's angular frequency $\Omega_H$ exceeds the wave's angular frequency $\omega$ as $m \Omega_H > \omega$, where $m$ is the magnetic quantum number. Energy extraction occurs in the ergoregion, leading to a superradiant instability that drains the BH's energy and angular momentum. This instability becomes significant at Bohr's radius, $r \sim (\alpha m_\phi)^{-1}$~\cite{Arvanitaki:2009fg}.

The system's dynamics are characterized by the fine structure constant $\alpha = r_g m_\phi$, such that a Bose-Einstein condensate can form around the BH. For astrophysical BHs (ApBHs), with masses $M_{\rm BH} \approx (1-100)\,M_\odot$, the instability becomes relevant for boson masses $m_\phi \sim (10^{-11}\mathrm{-}10^{-13})$\,eV. In the case of supermassive BHs (SMBHs), with masses $M_{\rm BH} \approx (10^6-10^9)\,M_\odot$, the corresponding boson mass is $m_\phi \sim (10^{-17}\mathrm{-}10^{-20})$\,eV~\cite{Brito:2014wla}. These mass ranges are not predicted in the Standard Model (SM), making superradiance a valuable tool for probing the BH surroundings~\cite{Khodadi:2021mct} and physics beyond the SM, including the axiverse scenario~\cite{Arvanitaki:2009fg, Arvanitaki:2010sy, Visinelli:2018utg} and theories of modified gravity~\cite{Khodadi:2020cht, Rahmani:2020vvv, Jha:2022tdl}. The observation of highly-spinning BHs excludes boson fields within the mass range $\sim r_g^{-1}$. The bosonic fields involved can have spin-0~\cite{Arvanitaki:2014wva, Stott:2018opm}, spin-1~\cite{Baryakhtar:2017ngi, Siemonsen:2022ivj}, or spin-2~\cite{Brito:2020lup}, with each spin exhibiting distinct instability timescales. As long as the superradiance condition holds, energy and angular momentum are extracted from the BH. Once this condition ceases, the bosonic cloud co-rotates with the BH. Self-interactions within the bosonic cloud, dominated by quartic terms, significantly alter the superradiant instability by enabling efficient energy transfer to radiative modes. This energy redistribution drives the system toward a quasi-equilibrium configuration at reduced occupation numbers, suppressing further cloud growth~\cite{Gruzinov:2016hcq, Baryakhtar:2020gao, Omiya:2022gwu, Witte:2024drg}. As a result, constraints based on maximal cloud amplitudes are relaxed.

Monochromatic GWs generated by transitions between energy levels or bosonic annihilations~\cite{Yoshino:2013ofa, Arvanitaki:2014wva, Stott:2018opm, Brito:2015oca, Baryakhtar:2017ngi, Brito:2017zvb, Siemonsen:2019ebd, Brito:2020lup, Zhu:2020tht, LIGOScientific:2021rnv, Siemonsen:2022ivj} could still be observable with next-generation GW observatories~\cite{LIGOScientific:2014pky, Amaro-Seoane:2012vvq, Punturo:2010zz, Maggiore:2019uih}, provided the suppression is not complete. Superradiant evolution could also be constrained by near-future observations with the Event Horizon Telescope~\cite{Roy:2019esk, Roy:2021uye, Chen:2022nbb, Ayzenberg:2023hfw}. However, these predictions rely on a relatively clean BH environment. Astrophysical surroundings, such as a dense plasma, can further modify the instability, either enhancing or quenching it depending on the coupling strength and modeling accuracy~\cite{Spieksma:2023vwl, Dima:2020rzg}. Plasma effects may induce new instabilities~\cite{Pani:2013hpa, Conlon:2017hhi}, although competing suppression mechanisms exist~\cite{Dima:2020rzg, Blas:2020kaa}.

Besides the Cosmic Microwave Background (CMB), the hot Big Bang model predicts the existence of a Cosmic Neutrino Background (C$\nu$B), which decoupled around one second after the Big Bang. Relic C$\nu$B neutrinos have energies in the range $(10^{-6}\textrm{--}10^{-4})$\,eV, making them difficult to detect. These neutrinos decoupled from the primordial plasma at a temperature of about 1\,MeV, long before CMB photons, which decoupled when $z_{\rm CMB} \approx 1100$ at $T_{\rm CMB} \approx 0.26$\,eV. Today, the temperature of relic neutrinos is $T_\nu = (4/11)^{1/3} T_{\rm CMB}/(1+z_{\rm CMB}) \approx 0.168$\,meV, with a corresponding number density per species $n_{\nu,0} \approx 56{\rm\,cm^{-3}}$ and the total density across the three active species $n_{\nu, {\rm tot}} \approx 336{\rm\,cm^{-3}}$~\cite{Dolgov:1997mb, Mangano:2005cc, Bauer:2022lri}. For Dirac neutrinos, $n_\nu = n_{\bar\nu} \approx n_{\nu,0}$, whereas for Majorana neutrinos, $n_\nu \approx 2n_{\nu,0}$. Anisotropies in the C$\nu$B may emerge due to finite neutrino masses~\cite{Lisanti:2014pqa}.

Relic neutrinos carry information on the cosmology prior CMB. Although their direct detection is challenging, the Princeton Tritium Observatory for Light, Early-Universe, Massive-Neutrino Yield (PTOLEMY) experiment~\cite{Betts:2013uya, PTOLEMY:2018jst, PTOLEMY:2019hkd} aims to observe these relic neutrinos through the inverse beta decay of tritium~\cite{Weinberg:1962zza, Cocco:2007za, Long:2014zva, Lisanti:2014pqa, Arteaga:2017zxg, Martinez-Mirave:2024dmw, Kim:2025xum}. Additionally, indirect methods, such as inelastic scattering with spin-polarized matter~\cite{Arvanitaki:2024taq} or screening of long-range forces~\cite{Smirnov:2019cae, Chauhan:2024qew}, may also be used to explore the C$\nu$B. Interactions between light bosonic DM and solar neutrinos could modify neutrino flavor oscillations~\cite{Berlin:2016woy, Krnjaic:2017zlz, Liao:2018byh, Capozzi:2018bps, Arguelles:2019xgp, Dev:2020kgz, Huang:2022wmz, Plestid:2024kyy}, although experiments like the Sudbury Neutrino Observatory (SNO)~\cite{SNO:2011hxd} and Super-Kamiokande (SuperK)~\cite{Super-Kamiokande:2001ljr} have not yet observed such effects. Upcoming experiments, including the Deep Underground Neutrino Experiment (DUNE)~\cite{DUNE:2015lol} and the Jiangmen Underground Neutrino Observatory (JUNO)~\cite{JUNO:2015zny}, will probe these interactions further~\cite{Krnjaic:2017zlz, Dev:2020kgz}. Boson-mediated neutrino interactions may also lead to astrophysical and cosmological consequences, such as influencing neutrino oscillations in supernovae~\cite{Duan:2010bg, Chakraborty:2016yeg}, delaying neutrino free-streaming in the early Universe~\cite{Cyr-Racine:2013jua, Archidiacono:2013dua, Lancaster:2017ksf, Oldengott:2017fhy, Kreisch:2019yzn, Visinelli:2024wyw}, and increasing the effective number of relativistic species~\cite{Wong:2002fa, Mangano:2006ar, Blinov:2019gcj, Brinckmann:2020bcn}.

In this paper, we investigate the impact of neutrinos on boson superradiance around spinning BHs, assuming a generation-independent Yukawa interaction between the scalar field and active neutrinos. The corrections to the scalar field's quartic self-interaction induced by both virtual and thermal neutrinos in the C$\nu$B impose constraints on the Yukawa coupling, based on superradiance conditions. These results are independent of whether the bosonic field is the DM, making them more robust than current constraints from cosmology, astrophysics, and laboratory experiments. Our findings complement existing searches by placing upper bounds on the boson-neutrino Yukawa coupling for the particular values of $m_\phi$ where superradiance is triggered. We use natural units with $\hbar = c = 1$.
    
\section{Superradiance driven by scalar fields}

A superradiant scalar field around a BH rotating with spin $a$ is described by a solution to the Klein-Gordon equation on the Kerr metric, with the boundary conditions specifying the incoming waves at the event horizon~\cite{Brill:1972xj}. For each eigenmode, the real component of the angular frequency is $\omega_{\bar n}\approx m_\phi$, where $\bar n >0$ is the principal quantum number, while the imaginary part characterizing the instability mode is expressed in terms of the azimuthal quantum number $l < \bar n$ and the positive integer $n = \bar n - l - 1 \geq 0$ as~\cite{Starobinskil:1974nkd, Page:1976df, Detweiler:1980uk}
\begin{equation}
    \begin{split}
    \Gamma_{nlm} &= 2m_\phi r_+(m\Omega_H-m_\phi)\alpha^{4l+4}\mathcal{A}_{nl}\,\mathcal{X}_{nl}\,,\label{sp4}\\
    \mathcal{A}_{nl} &= \frac{2^{4l+2}\,(2l+n+1)!}{(l+n+1)^{2l+4}\,n!}\,\left(\frac{l!}{(2l)!(2l+1)!}\right)^2\,,\\ 
    \mathcal{X}_{lm} &= \prod_{j=1}^l\,\left[j^2(1-a_*^2)+4r_+^2(m\Omega_H - m_\phi)^2\right]\,,
    \end{split}    
\end{equation}
where $a_*\equiv a/r_g$ is the spin parameter and $r_+ = r_g(1+\sqrt{1-a_*^2})$ is the outer physical BH event horizon, with angular velocity
\begin{equation}
    \Omega_H = \frac{1}{2r_g}\frac{a_*}{1+\sqrt{1-a_*^2}}\,.
    \label{sp5}
\end{equation}
The superradiance condition is determined by the sign of $\Gamma_{nlm}$. As long as $m\Omega_H>m_\phi$, the amplitude of the wave solution exhibits an exponential growth with the occupation number of particles in the bosonic cloud increasing as $\dot N_{nlm} = \Gamma_{nlm}\,N_{nlm}$. The bosons need not be initially present around the BH or be part of the DM. It can arise from quantum fluctuations and grow exponentially via superradiance, continuing as long as the superradiant timescale is shorter than the characteristic timescale over which the angular momentum of the BH changes due to other physical processes. ApBHs cannot accrete faster than the Eddington mass accretion rate, where the outward radiation pressure balances the inward pull of gravity. This limitation leads to the Salpeter timescale, $\tau_{\rm Sal} \approx 4.5 \times 10^7$\,yr~\cite{Salpeter:1964kb, Shakura:1972te}. For supermassive BHs, we set $\tau_{\rm BH} = 10^9$\,yr following Ref.~\cite{Davoudiasl:2019nlo}. These considerations define an exclusion region on the Regge plane $(M_{\rm BH}, a_*)$, which is compared to measurements of BH spin and mass. When the superradiance condition no longer applies, the maximum number of bosons in a given level is~\cite{Arvanitaki:2014wva}
\begin{equation}
    N_{\mathrm{max}} = \frac{G M_{\rm BH}^2\Delta a_*}{m} \approx \frac{10^{76}}{m}\,\left(\frac{M_{\rm BH}}{10\,M_\odot}\right)^2\,\left(\frac{\Delta a_*}{0.1}\right),
\end{equation}
where $\Delta a_*$ represents the change in BH spin from its initial to final state. The imaginary component of the frequency, $\Gamma_{nlm}$, decreases with increasing $l$, so the $l=1$ mode grows most rapidly; $\Gamma_{nlm}$ peaks at $m = l$ and diminishes significantly for $m < l$. Additionally, as $l$ increases, the quantum number $n$ also rises for the fastest-growing mode~\cite{Yoshino:2013ofa}.

Superradiant amplification can be suppressed by self-interactions within the scalar field~\cite{Yoshino:2012kn, Fukuda:2019ewf, Baryakhtar:2020gao}, and level-mixing of different bosonic states can even preclude superradiance altogether~\cite{Arvanitaki:2010sy, Yoshino:2012kn, Yoshino:2015nsa}. Even a small quartic self-coupling $\lambda$ can significantly alter the dynamics by enabling energy transfer between bound states and the escape of scalar radiation to infinity~\cite{Baryakhtar:2020gao, Witte:2024drg}. This effectively suppresses the superradiant instability and, in some regimes, can preclude the formation of a dominant cloud or delay its growth on astrophysically relevant timescales.

\section{Cosmic neutrinos and light scalars}

The Lagrangian for a light scalar field $\phi$ coupled to a relic active Dirac neutrino species $\nu_\alpha$ is
\begin{equation}
    \label{eq:Lagrangian}
    \mathcal{L} \supset \frac{1}{2}\partial_\mu\phi\partial^\mu \phi-\frac{1}{2}m^2_\phi\phi^2 - m_{\alpha\beta}\bar{\nu}_\alpha\nu_\beta-y_{\alpha\beta}\phi\bar{\nu}_\alpha\nu_\beta\,,
\end{equation}
where the last term represents the scalar-relic neutrino interaction and $m_{\alpha\beta}$ is the neutrino mass matrix. The species-dependent Yukawa coupling $y_{\alpha\beta}$, which parameterizes the non-standard neutrino interaction, can be induced in low-energy observables via an effective four-fermion interaction~\cite{Wolfenstein:1977ue, Guzzo:1991hi, Dvali:2016uhn, Farzan:2017xzy, Funcke:2019grs}. Here, we assume it reduces to a universal coupling $y_{\phi\nu}$ between any neutrino generation and the scalar field.

In a non-supersymmetric (SUSY) theory, radiative corrections to the scalar potential from fermion loops generate an effective quartic term even in the absence of a tree-level quartic coupling. The Coleman-Weinberg mechanism yields an effective potential $V_{\rm CW} \propto y_{\phi\nu}^4\phi^4\ln\phi^2$, with an associated coupling~\cite{Coleman:1973jx}
\begin{equation}
    \lambda^{(0)} \sim \frac{y_{\phi\nu}^4}{16\pi^2}\,\ln\left(\frac{m_\phi^2}{m_\nu^2}\right)\,.     
\end{equation}
In SUSY, by contrast, quartic terms cancel at zero temperature, and radiative corrections to the scalar potential are suppressed. However, thermal effects break SUSY, allowing the scalar field to acquire both a thermal mass and a self-interaction through its Yukawa coupling to fermions~\cite{Nadkarni:1988fh, Baier:1991dy, Thoma:1994yw}.

We focus on thermal corrections from the C$\nu$B when $m_\phi \ll T_\nu, m_\nu$. The mass correction reads\footnote{See Eq.~(D10) in Ref.~\cite{Babu:2019iml} for the general expression if $m_\phi\!\neq\!0$.}
\begin{equation}
    \label{eq:masstermcontribution}
    \Delta m_\phi^2 = \frac{y_{\phi\nu}^2}{\pi^2}\,\int_{m_\nu}^{+\infty}{\rm d}\varepsilon\,\sqrt{\varepsilon^2-m_\nu^2}\,f_\nu(\varepsilon)\,,
\end{equation}
where $f_\nu(\varepsilon)$ is the Fermi-Dirac distribution at energy $\varepsilon$, assuming vanishing chemical potential. The inverted hierarchy with $m_1=m_2=50$\,meV, $m_3$=10\,meV yields
\begin{equation}
    \label{eq:deltam2ModelII}
    \Delta m_\phi^2 = 1.2\times 10^{-10}{\rm\,eV^2}\,y_{\phi\nu}^2\,,
\end{equation}
leading to an effective scalar mass $m_{\phi,{\rm eff}}^2 = m_\phi^2 + \Delta m_\phi^2$. Results would not change appreciably when other scenarios such as normal mass ordering or a massless neutrino eigenstate are taken into account.\footnote{The neutrino mass would also receive corrections from boson interactions~\cite{Babu:2019iml, Nieves:2021oll}.}

A quartic Lagrangian term is induced through a loop that involves the Yukawa coupling in Eq.~\eqref{eq:Lagrangian}, as
\begin{equation}
    \label{eq:fermionloop}
    \mathcal{L}_{\rm int} = \frac{1}{4!}\lambda\phi^4\,,
\end{equation}
where the quartic coupling $\lambda = \lambda^{(0)} +\Delta\lambda^{(T)}$ is the sum of the vacuum contribution $\lambda^{(0)}$ and the thermal correction which, in the limit $m_\nu\gg T_\nu$, reads
\begin{equation}
    \Delta \lambda^{(T)} \sim y_{\phi\nu}^4 \frac{n_{\nu, {\rm tot}}}{m_\nu^3}\,,
\end{equation}
see Appendix~\ref{sec:appendix} for details. Here, $n_{\nu, {\rm tot}}$ is the C$\nu$B number density which has the cosmological value $n_{\nu, {\rm tot}} \approx 336{\rm\,cm^{-3}}$. The self-interaction term triggers non-linear dissipation effects at values of $y_{\phi\nu}$ where the mass correction in Eq.~\eqref{eq:deltam2ModelII} can be safely neglected. The efficiency of energy extraction is then limited by the non-linear dynamics of Eq.~\eqref{eq:fermionloop}.

\section{Methods}

The Yukawa interactions in Eqs.~\eqref{eq:Lagrangian} and~\eqref{eq:fermionloop} lead to trajectories on the Regge plane $(M_{\rm BH}, a_*)$. These are compared with data for the mass and spin of known BHs available in the literature. For the analysis, we separately consider two observed ApBH: GRS 1716-249~\cite{Fishbach:2021xqi, Tao:2019yhu} and Cygnus X-1~\cite{Miller-Jones:2021plh, Zdziarski:2024zfg}, along with five SMBHs, Mrk 79~\cite{Peterson:2004nu, Gallo:2010mm}, NGC 4051~\cite{Peterson:2004nu, 2009ApJ...702.1353D, Patrick:2012ua}, MCG-6-30-15~\cite{McHardy:2005ut, Brenneman:2006hw}, Ark 120~\cite{Peterson:2004nu, Walton:2012aw}, and M87$^*$~\cite{EventHorizonTelescope:2019ggy, Tamburini:2019vrf}. The properties of these BHs are summarized in Table~\ref{table:BHs}. These BHs are chosen due to their high observed spin, which makes them ideal targets for exploring a larger portion of the Regge plane. For M87$^\star$, we adopt the spin value $a = (0.9\pm0.05)$~\cite{Tamburini:2019vrf}, which constrains the existence of boson fields with masses around $\sim 10^{-21}$\,eV~\cite{Davoudiasl:2019nlo}. Although M87$^*$ is believed to be highly spinning, measurements of its spin have not yet reached a consensus. The constraints for NGC 4051 are affected by modeling issues, resulting in poorly constrained or extreme disk parameters~\cite{Patrick:2012ua}, such as high emissivity and the near-maximal spin in Table~\ref{table:BHs}. In contrast, while unmodeled emissions could lower the spin inferred for MCG-6-30-15, all models consistently suggest a near-maximal spin, strongly indicating that the data favor a rapidly spinning BH~\cite{Brenneman:2006hw}.
\begin{table}
    \begin{center}
    \renewcommand{\arraystretch}{1.3}
    \begin{tabular}{|l@{\hspace{0.1 cm}}|l@{\hspace{0.1 cm}}|l@{\hspace{0.1 cm}}|l@{\hspace{0.1 cm}}|}
        \hline
       \text{ApBH} & $M_{\rm BH}\,[M_{\odot}]$ & $a_*$ & Spin CL\\
        \hline
GRS 1716-249 & $6.45 \pm 1.55$ & $\geq 0.92$ & 90\% \\ 
Cygnus X-1 & $21.2 \pm 2.2$ & $0.92^{+0.05}_{-0.07}$ & 99.7\% \\
\hline
\text{SMBH} & $M_{\rm BH}\,[10^6\,M_{\odot}]$ & $a_*$ & Spin CL\\
\hline
NGC 4051 & $1.91 \pm 0.78$ & $\geq 0.99$ & 90\% \\
MCG-6-30-15 & $2.90^{+1.80}_{-1.60}$ & $0.989^{+0.009}_{-0.002}$ & 90\%\\
Mrk 79   & $52.40 \pm 14.40$ & $0.70^{+0.1}_{-0.1}$ & 90\% \\
Ark 120 & $150.0 \pm 19.0$ & $0.64^{+0.19}_{-0.11}$ & 90\% \\
M87* & $6500.0 \pm 700.0$ & $0.9^{+0.05}_{-0.05}$ & 95\%\\
    \hline
    \end{tabular}
    \caption{Mass and dimensionless spin parameters measurements used in the analysis. Mass uncertainties are quoted at 1$\sigma$. See text for references.}
    \label{table:BHs}
\end{center}
\end{table}
We perform a least-squares fit based on a maximum likelihood estimator that incorporates the measured spin and mass of each BH. In our analysis, we vary the scalar field mass and Yukawa coupling while assuming Gaussian errors, following the methodology outlined in Refs.~\cite{Fernandez:2019qbj, Cheng:2022jsw}. Given that BH spin measurements often have asymmetric uncertainties around their central values, we model the likelihood using a split normal distribution, which reduces to a normal distribution when the errors are symmetric~\cite{doi:10.1080/01621459.1998.10474117}. The black hole spin is evolved along with the occupation numbers of two bound states, $|n,\ell,m\rangle = |211\rangle$ and $|322\rangle$, following the framework developed in Refs.~\cite{Baryakhtar:2020gao, Witte:2024drg}.

\section{Results}

\begin{figure}[ht]
    \centering
    \includegraphics[width=\linewidth,keepaspectratio]{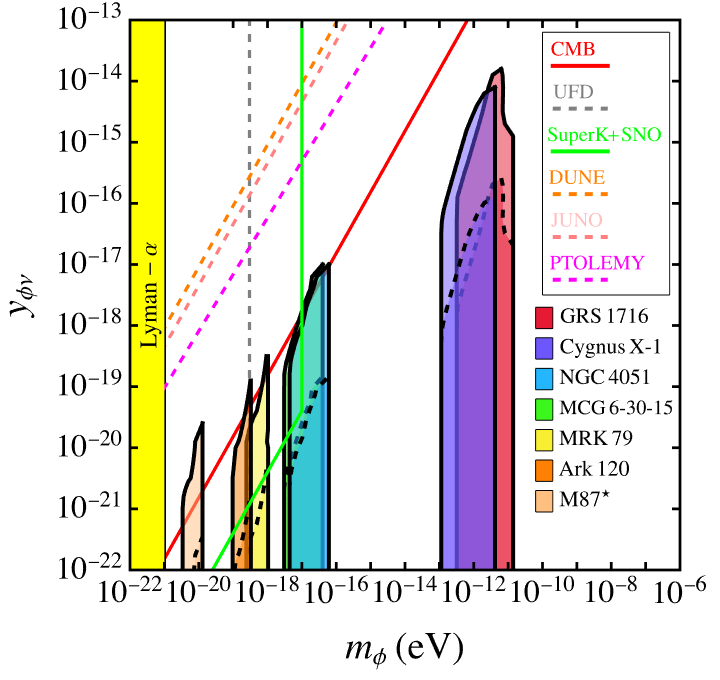}
    \caption{Constraints on the scalar-neutrino Yukawa coupling $y_{\phi\nu}$ as a function of the boson mass $m_\phi$. The shaded regions are excluded by the superradiant instability triggered by the self-coupling $\lambda$ in Eq.~\eqref{eq:fermionloop}, computed either from vacuum fluctuations (dashed contours, shaded regions) or including C$\nu$B thermal corrections (solid contours, light shaded regions), assuming inverted neutrino mass ordering. Also shown are constraints from astrophysical, cosmological, and laboratory observations; see the text for details.}
    \label{fig:constraints}
\end{figure}

Figure~\ref{fig:constraints} shows the constraints on the Yukawa coupling $y_{\phi\nu}$ including all superradiant unstable modes within the BH accretion timescale. We present results assuming either (i) vacuum-induced quartic coupling $\lambda^{(0)}$ (dashed contours, shaded regions) or (ii) suppressed $\lambda^{(0)}$ due to SUSY cancellations and dominant thermal corrections as in Eq.~\eqref{eq:fermionloop} (solid contours, light shaded regions). In both cases, non-linear effects quench the instability, yielding upper bounds on $y_{\phi\nu}$. Because superradiance limits $\lambda$, thermal-induced constraints are generally stronger. Table~\ref{table:results} summarizes the 1$\sigma$ confidence level (CL) bounds on the boson mass ($m_\phi$/eV) for the BHs in Table~\ref{table:BHs}.
\begin{table}[!ht]
    \begin{center}
    \renewcommand{\arraystretch}{1.3}
    \begin{tabular}{|l@{\hspace{0.1 cm}}|c@{\hspace{0.1 cm}}|}
    \hline
    \text{BH} &Bound on $m_\phi/{\rm\,eV}$ \\
    \hline
GRS 1716-249 & $[0.03,1.4]\times10^{-11}$\\
Cygnus X-1 & $[0.1, 4.1]\times10^{-12}$\\
\hline
NGC 4051 & $[0.4,6.1]\times10^{-17}$\\
MCG-6-30-15 & $[0.3,4.0]\times10^{-17}$\\
Mrk 79   & $[0.2, 1.0]\times10^{-18}$\\
Ark 120 & $[0.9, 3.3]\times 10^{-19}$\\
M87* & $[0.3, 1.3]\times 10^{-20}$\\
 \hline
    \end{tabular}
    \caption{Results for the least square method with the maximum likelihood estimator described in the text and the BHs in Table~\ref{table:BHs}. Bounds are quoted at 1$\sigma$.}
    \label{table:results}
\end{center}
\end{table}

Constraints on scalar-neutrino interactions arise from astrophysical, cosmological, and laboratory observations, often assuming the scalar field constitutes DM. Ultralight scalar DM suppresses small-scale cosmic structures via quantum pressure, with Lyman-$\alpha$ observations setting $m_\phi \gtrsim 10^{-21}$ eV~\cite{Kobayashi:2017jcf} (yellow vertical band). Additional constraints follow from neutrino mass corrections at recombination, predicted by the Lagrangian in Eq.~\eqref{eq:Lagrangian}, as $\delta m_i \sim y_{\phi\nu}\,\langle|\phi|\rangle$, with $\langle|\phi|\rangle \sim \sqrt{2\rho_{\rm DM}}/m_\phi$. Comparing this with CMB limits on the sum of neutrino masses in $\Lambda$CDM+$m_\nu$ model~\cite{Vagnozzi:2017ovm, Planck:2018vyg, Jiang:2024viw} yields the bound (red line in Fig.~\ref{fig:constraints})
\begin{equation}
    y_{\phi\nu} \lesssim 1.5 \times 10^{-23}\,\left(\frac{m_\phi}{10^{-22}{\rm\,eV}}\right)\,\quad \hbox{(CMB)}\,.
\end{equation}
This is comparable to constraints from neutrinoless double beta decay at recombination~\cite{Huang:2021kam}. In astrophysics, dynamical heating of stellar orbits in ultrafaint dwarf (UFD) galaxies sets a stringent limit on the DM particle mass, $m_\phi \gtrsim 3 \times 10^{-19}$\,eV at 99\% CL~\cite{Dalal:2022rmp}. Moreover, scalar-neutrino interactions can alter neutrino oscillation parameters, affecting mass splittings and mixing angles by changing oscillation amplitudes and frequencies due to the time-varying scalar DM field. The absence of periodic fluctuations in solar neutrino data from SNO and SuperK constrains the scalar-neutrino coupling to $y_{\phi\nu} \sim 4\times10^{-21}\,(m_\phi/10^{-18}\,\mathrm{eV})$~\cite{Berlin:2016woy} (green line in Fig.~\ref{fig:constraints}). Observational durations $t_{\rm obs} \gtrsim 1$\,min imply a mass cutoff $m_\phi \sim 2\pi / t_{\rm obs} \lesssim 7\times 10^{-17}$\,eV. Future experiments could further tighten these constraints or detect scalar-neutrino coupling through time-modulated neutrino oscillations~\cite{Krnjaic:2017zlz, Dev:2020kgz}, with sensitivity to periods down to tens of milliseconds for DUNE (orange) and shorter for JUNO (pink). The PTOLEMY sensitivity (magenta) is estimated from Ref.~\cite{PTOLEMY:2019hkd}. This approach highlights connections between scalar fields, neutrinos, and cosmic phenomena, calling for further investigation in future experiments.

\section{Discussions and conclusions}

The results in Fig.~\ref{fig:constraints} derived from thermal effects assume a relic neutrino density of $n_{\nu, {\rm tot}} \approx 336{\rm\,cm^{-3}}$, consistent with standard C$\nu$B expectations~\cite{Dolgov:1997mb, Mangano:2005cc, Bauer:2022lri}. Several scenarios can modify this assumption and shift the excluded region in $y_{\phi\nu}$. For instance, our analysis assumes a homogeneous C$\nu$B when computing the scalar effective mass. However, gravitational clustering can enhance the local neutrino density near BHs. Massive neutrinos accumulate within the gravitational influence radius $r_{\rm infl} = 2r_g/v_\nu^2$, where $v_\nu$ is the neutrino thermal velocity. For relic neutrinos of mass $m_\nu = 50$\,meV, the thermal velocity is $v_\nu\sim 10^{-3}$, with the corresponding $r_{\rm infl} \sim 10^6\,r_g$. Inside this region, the neutrino number density is enhanced and follows the radial profile~\cite{Perlick:2015vta}
\begin{equation}
    n_{\rm spike}(r) =n_{\nu, {\rm tot}}\,(r_{\rm infl}/r)^{3/2}\,.
\end{equation}
If superradiance is driven by the effective mass term in Eq.~\eqref{eq:masstermcontribution}, then spatial variations in the neutrino density induce a radial dependence in the scalar effective mass. This effect can suppress the instability by removing bound state solutions~\cite{Dima:2020rzg, Wang:2022hra, Cannizzaro:2022xyw}. A neutrino spike increases the local density and weakens the superradiance bounds by about one order of magnitude, since $y_{\phi\nu} \to y_{\phi\nu} (n_{\nu, {\rm tot}} / n_{\rm spike})^{1/4} \sim y_{\phi\nu} (T_\nu/ \alpha m_\nu)^{3/8}$. Another non-linear effect is $\phi$–$\nu$ scatterings, which remains subdominant to mode-mode depletion across the relevant parameter space and does not quench the superradiant instability. While the growing cloud backreacts on the neutrino background, the resulting dissipation remains inefficient due to the suppressed cross section from heavy fermion exchange. Similar nonlinear backreaction effects have been shown to limit plasma-driven instabilities~\cite{Cardoso:2020nst, Blas:2020kaa, Cannizzaro:2023ltu}.

An additional galactic enhancement in $n_{\nu, {\rm tot}}$, due to gravitational clustering, is also expected, the exact value depending on the neutrino mass~\cite{Zhang:2017ljh, Mertsch:2019qjv, Holm:2024zpr}. Interactions with cosmic rays also modify the energy distribution of the C$\nu$B~\cite{Ciscar-Monsalvatje:2024tvm}. The late decay of hidden particles into neutrinos~\cite{Chacko:2018uke} or primordial BHs~\cite{Lunardini:2019zob} could alter the relation between the CMB and C$\nu$B temperatures, affecting the estimate of $n_{\nu,0}$. Other effects that could change the current background include neutrino clustering~\cite{Ringwald:2004np, Mertsch:2019qjv}, decay~\cite{Bernal:2021ylz, Carenza:2023qxh}, and screening~\cite{Visinelli:2024wyw}. Although the diffuse supernova neutrino background constitutes a very bright source, its number density $\approx 10^{-11}{\rm\,cm}^{-3}$~\cite{DeGouvea:2020ang,Ando:2023fcc,MacDonald:2024vtw} is too low to significantly alter our results.

In addition to neutrinos, other SM fermions, such as electrons, can modify the effective mass of the boson field through thermal corrections. This effect depends on the electron number density, $n_e$, in the background medium, such as the BH's accretion disk, contributing to the boson mass as
\begin{equation}
    \Delta m_\phi^2 = 3\times 10^{-10}{\rm\,eV^2}\,y_{\phi e}^2\left(\frac{n_e}{10^{10}{\rm\,cm^{-3}}}\right),
\end{equation}
where $y_{\phi e}$ is the scalar-electron Yukawa coupling. Electron densities near BH horizons in thin or thick accretion disks are orders of magnitude higher than in the interstellar medium $n_e \lesssim 10^{-2}{\rm\,cm^{-3}}$~\cite{Dima:2020rzg}, reaching $n_e \sim 10^{19}{\rm\,cm^{-3}}\,(M_\odot/M_{\rm BH})$ in advection-dominated accretion flows~\cite{Narayan:1994is}. This could significantly affect the boson mass, even for small $y_{\phi e}$. This potential effect may require a dedicated study of individual BHs, where spectral and timing analyses of X-ray observations allow for the investigation of the accretion disk. A detailed study of BHs observed as a soft X-ray transients like GRS 1124-683~\cite{Esin:1997he} could explore these effects via spectral and timing analyses. BH superradiance constraints leave the boson mass range $(10^{-17}\textrm{--}10^{-13})$\,eV unexplored due to the absence of evidence for spinning intermediate-mass BHs. Novel detection strategies, such as those involving quasi-periodic eruptions~\cite{2020ARA&A..58..257G, Zhou:2024vwj}, may address this gap in the near future.

In this paper, we have examined superradiant interaction rates for a light scalar field around highly spinning BHs, incorporating interactions with the C$\nu$B via a Yukawa coupling, $y_{\phi\nu}$ that lead to thermal corrections of the scalar quartic coupling and induce superradiant instability. A least-square analysis that includes astrophysical and supermassive BHs leads to novel constraints on the Yukawa coupling as a function of the scalar field mass $m_\phi$. Assuming two degenerate neutrino masses with inverted ordering, we find that superradiance excludes couplings as large as $y_{\phi\nu} \sim 10^{-16}$ ($y_{\phi\nu} \sim 10^{-20}$) for ApBHs (SMBHs), the exact value depending on $m_\phi$. Our results are independent of whether the scalar field constitutes DM and complement existing and perspective bounds on the scalar-neutrino couplings.

\vspace{.3cm}
\begin{acknowledgments}
We thank Rudnei O.~Ramos and Richard Brito for reading a preliminary version of the draft and for their constructive comments. We also thank Yifan Chen and Sam Witte for helpful feedback that enriched this study, as well as Sunny Vagnozzi and Edoardo Vitagliano for stimulating discussions. This work is supported by INFN through the ``QGSKY'' Iniziativa Specifica project. L.V.\ also acknowledges support by the National Natural Science Foundation of China (NSFC) through the grant No.\ 12350610240 ``Astrophysical Axion Laboratories''. This publication is based upon work from the COST Actions ``COSMIC WISPers'' (CA21106) and ``Addressing observational tensions in cosmology with systematics and fundamental physics (CosmoVerse)'' (CA21136), both supported by COST (European Cooperation in Science and Technology).
\end{acknowledgments}

\appendix

\section{Induced Quartic Coupling from Fermion Box Diagram}
\label{sec:appendix}

\subsection{Vacuum Contribution}

\begin{figure}[thb]
    \centering
    \includegraphics[width=0.7\linewidth]{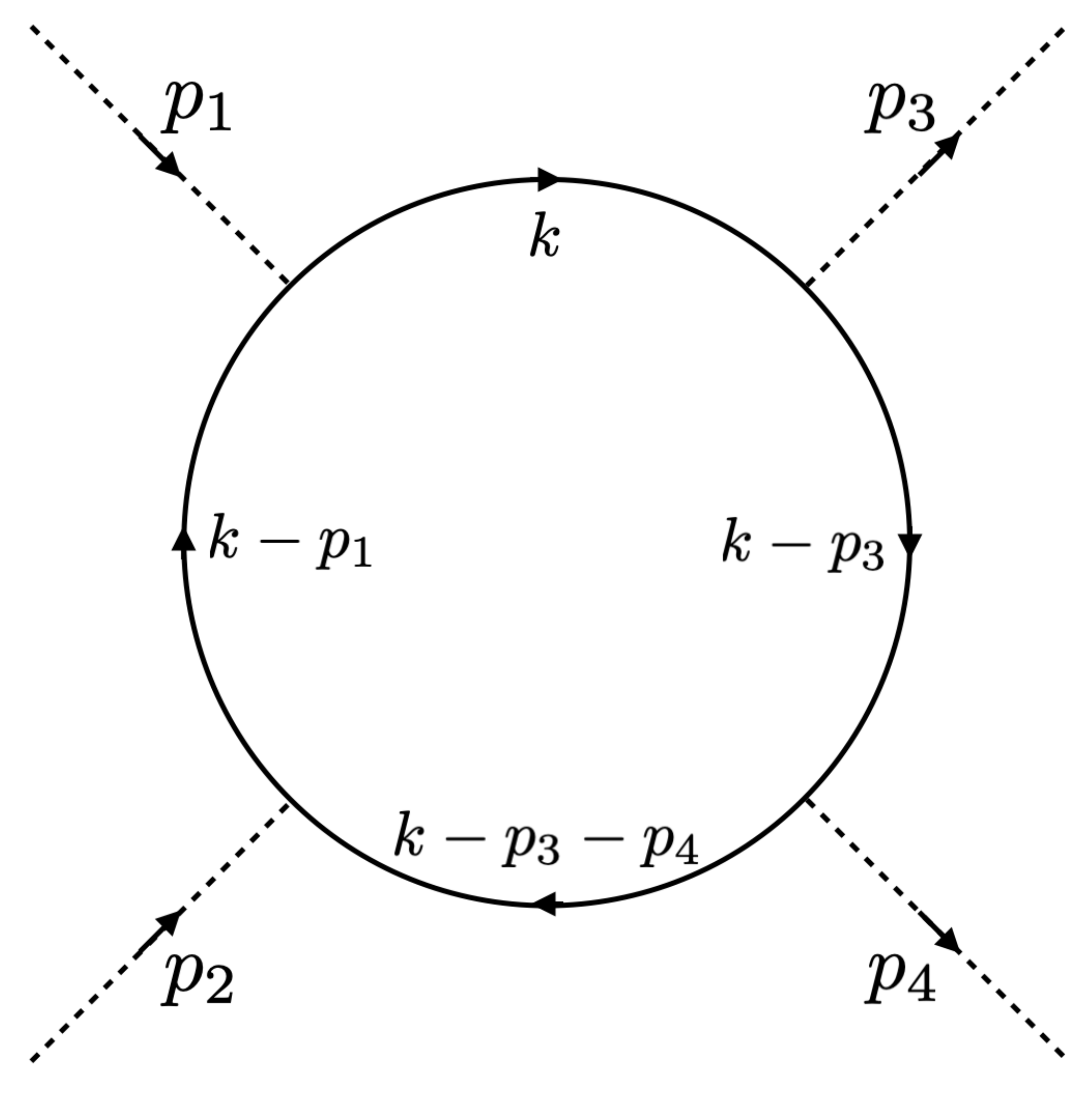}
    \caption{Thermal contribution to the quartic coupling.}
    \label{fig:quartic}
\end{figure}

Consider a real scalar field $\phi$ coupled to a Dirac neutrino $\nu$ via a Yukawa interaction:
\begin{equation}
    \mathcal{L}_{\text{int}} = y_{\phi\nu} \, \phi \, \bar{\nu} \nu \,.
\end{equation}
At one-loop level, a quartic self-interaction for $\phi$ is induced via a fermion box diagram.\footnote{For the computation, see e.g.\ Chapter~6 of Peskin and Schroeder's book~\cite{Peskin:1995ev}.} This diagram consists of four external scalar legs and a loop of neutrinos, as sketched in Fig.~\ref{fig:quartic}

Here, the quartic self-interaction mediates between two bosons undergoing momentum exchange as $p_1+p_2 \to p_3+p_4$. The corresponding amplitude is
\begin{equation}
\begin{split}
\mathcal{M} \;\sim\; &y_{\phi\nu}^4 \int \frac{{\rm d}^4 k}{(2\pi)^4} \, \text{Tr} \Bigg[
\frac{\not{k} + m_\nu}{k^2 - m_\nu^2}
\frac{\not{k} - \not{p}_1 + m_\nu}{(k - p_1)^2 - m_\nu^2}\\
&\times \frac{\not{k} - \not{p}_3 + \not{p}_4 + m_\nu}{(k - p_3 - p_4)^2 - m_\nu^2}
\frac{\not{k} - \not{p}_3 + m_\nu}{(k - p_3)^2 - m_\nu^2}
\Bigg] \,.
\end{split}
\end{equation}
In the low-energy limit, where the external momenta satisfy $|p_i| \ll m_\nu$, the integral is dominated by $k \sim m_\nu$, and one can expand in powers of $p_i / m_\nu$. At leading order, the box diagram reduces to the induced quartic coupling
\begin{equation}
    \label{eq:lambdavacuum}
    \boxed{
        \lambda^{(0)} \sim \frac{y_{\phi\nu}^4}{16\pi^2}\,\ln\left(\frac{m_\phi^2}{m_\nu^2}\right)\,.
    }
\end{equation}

\subsection{Thermal corrections to the boson mass and the quartic coupling term}
\label{sec:QFTmass}

The thermal correction can be computed using the Matsubara formalism. The correction is given schematically by
\begin{equation}
    \Delta \lambda^{(T)} \sim y_{\phi\nu}^4 T \sum_{n} \int \frac{{\rm d}^3 {\bf k}}{(2\pi)^3} \, \text{Tr}\left[ G(i\omega_n, {\bf k})^4 \right] \,,
\end{equation}
where the fermion propagator is
\begin{equation}
G(i\omega_n, {\bf k}) = \frac{i\omega_n \gamma^0 - {\bf \gamma}\cdot {\bf k} + m_\nu}{\omega_n^2 + E_k^2}, \quad E_k = \sqrt{{\bf k}^2 + m_\nu^2} \,.
\end{equation}
We first perform the Dirac trace to find a numerator of the order of $(\omega_n^2 + E_k^2)^2$. The amplitude simplifies to
\begin{equation}
\lambda_\phi(T)
\!\propto\!
y_{\phi\nu}^4 T \sum_{n}\!\int\!\frac{{\rm d}^3{\bf k}}{(2\pi)^3}
\frac{\omega_n^4 \!+\! 2\omega_n^2({\bf k}^2 \!-\! m_\nu^2) \!+\! ({\bf k}^2 \!+\! m_\nu^2)^2}
{(\omega_n^2 + E_k^2)^4}.
\end{equation}
The Matsubara‐sum techniques allow us to convert
\begin{equation}
\begin{split}
    T\sum_n \frac{1}{(\omega_n^2 + E_k^2)^m}
    \;\longrightarrow\;
    &\frac{1}{2E_k} \frac{(-1)^{m-1}}{(2E_k)^{2m-2}}
    \Big[1 - 2f_\nu(E_k)\Big]\\
    &\times(\text{polynomial in }E_k)\,.
\end{split}
\end{equation}
After performing the Matsubara sum, one finds
\begin{equation}
    \lambda_\phi(T) = \lambda^{(0)} + \Delta\lambda^{(T)}\,,
\end{equation}
where the vacuum piece reproduces the quartic coupling in Eq.~\eqref{eq:lambdavacuum} and the thermal correction comes from the $-2f_\nu(E_k)$ terms. Extracting the finite-temperature piece, the leading contribution is
\begin{equation}
\label{eq:lambdathermal}
\boxed{\Delta \lambda^{(T)} \sim y_{\phi\nu}^4 \int \frac{{\rm d}^3{\bf k}}{(2\pi)^3} \frac{f_\nu(E_k)}{E_k^3} \sim  y_{\phi\nu}^4 \frac{n_{\nu, {\rm tot}}}{m_\nu^3}\,,}
\end{equation}
where $f_\nu(E_k)$ is the Fermi-Dirac distribution and in the last step we have assumed $T_\nu \ll m_\nu$. The sum of Eqs.~\eqref{eq:lambdavacuum} and~\eqref{eq:lambdathermal} accounts for both the virtual contribution from the fermion box and the thermal effects from the ambient neutrino background.

\subsection{Implications of Supersymmetry on Scalar Self-Interactions}
\label{subsec:SUSY_effects}

Supersymmetry (SUSY) imposes stringent constraints on the structure of scalar potentials, with important implications for both the tree-level quartic couplings and their radiative corrections. In this subsection, we summarize how SUSY affects the emergence of scalar self-interactions and clarify why thermal contributions can remain sizeable even when vacuum effects are suppressed. In supersymmetric models, scalar potentials are determined by F-terms that arise from the superpotential $W(\Phi)$, and D-terms associated with gauge interactions and generating quartic couplings only if the scalar is charged under a gauge group. As a result, in models where the scalar is a gauge singlet and the superpotential is minimal, there is no tree-level quartic interaction. Quartic terms from F-terms are constrained by the holomorphic structure of $W(\Phi)$, while D-terms vanish for uncharged fields. Therefore, scalar quartic couplings $\sim \lambda \phi^4$ are often absent or non-generic in minimal SUSY setups.

Radiative corrections to the scalar potential in SUSY are also constrained. In the limit of exact SUSY, bosonic and fermionic loop contributions cancel due to equal masses and opposite statistics:
\begin{equation}
    \delta V_{\text{1-loop}} \propto \sum_i (-1)^{F_i} n_i m_i^4 \log\left( m_i^2/\mu^2 \right)\,,
\end{equation}
where $F_i = 0$ for bosons and $F_i = 1$ for fermions, $n_i$ counts the degrees of freedom, and $m_i$ are the field-dependent masses. This cancellation ensures that radiative quartic corrections vanish in the vacuum while flat directions in the scalar potential are preserved. In realistic models, SUSY must be broken. Soft SUSY-breaking terms introduce mass splittings between superpartners, reintroducing loop corrections. However, these corrections are still softened compared to generic non-SUSY theories. For instance, in the MSSM, the Higgs quartic coupling is determined at tree level by gauge couplings (via D-terms), and only receives logarithmic corrections from top/stop loops, which are essential to accommodate the observed Higgs mass.

At finite temperature, supersymmetry is explicitly broken by the thermal environment. The bosonic and fermionic thermal distribution differ, breaking the degeneracy and leading to incomplete cancellations in loop diagrams. For this, thermal loop corrections to scalar self-interactions remain non-zero, even in supersymmetric theories. These corrections can induce effective quartic couplings (e.g., from fermion box diagrams), despite vanishing vacuum contributions. This mechanism is particularly relevant in the present work, since a neutrino background at finite temperature can generate a thermal quartic term for a scalar field through neutrino box diagrams, even in models where vacuum contributions would cancel due to SUSY. This behavior is consistent with results from early-universe cosmology, where finite-temperature effects induce scalar thermal masses and interactions even in supersymmetric models.

\bibliographystyle{apsrev4-1}
\bibliography{superradiance.bib}

\end{document}